\documentclass[aps,a4paper,twocolumn]{revtex4}
\usepackage{amsmath}
\usepackage{graphicx}
\usepackage{subfigure}
\usepackage[usenames,dvipsnames]{color}
\definecolor{darkblue}{RGB}{0,0,196}
\usepackage[colorlinks=true,linkcolor=darkblue,citecolor=darkblue,urlcolor=darkblue]{hyperref}
\usepackage{setspace}
\usepackage{url}
\usepackage{hyperref}
\usepackage{ulem}
\usepackage{xcolor}
\hypersetup{
  colorlinks   = true, 
  urlcolor     = red, 
  linkcolor    = blue, 
  citecolor   = blue 
}
\usepackage{graphicx}
\usepackage{amsmath,bbm}
\usepackage{amssymb,bm}
\usepackage{yfonts}
\usepackage{comment}

\usepackage[usenames,dvipsnames]{color}
\definecolor{darkblue}{RGB}{0,0,196}
\usepackage[colorlinks=true,linkcolor=darkblue,citecolor=darkblue,urlcolor=darkblue]{hyperref}
\begin{document}
\title{Vorticity-induced modifications of chemical freeze-out in heavy-ion collisions}

\author{Nandita Padhan$^{1}$} 
\author{Kshitish Kumar Pradhan$^{2}$} 
\author{Arghya Chatterjee$^{1}$\footnote{achatterjee.phy@nitdgp.ac.in}} 
\author{Raghunath Sahoo$^{2}$\footnote{Corresponding Author Email: Raghunath.Sahoo@cern.ch}} 
\affiliation{$^{1}$Department of Physics, National Institute of Technology Durgapur, Durgapur 713209, West Bengal, India}
\affiliation{$^{2}$Department of Physics, Indian Institute of Technology Indore, Simrol, Indore 453552, India}

\begin{abstract}
We investigate the influence of global rotation on the chemical freeze-out parameters in ultra-relativistic heavy-ion collisions. Within the framework of the hadron resonance gas (HRG) model, the freeze-out parameters are determined using commonly employed freeze-out criteria, namely the fixed energy per particle and the scaled entropy density, extended here to include rotational effects. 
We find that the presence of rotation leads to a systematic shift of the chemical freeze-out curve toward lower temperatures in the $T\text{--}\mu_B$ phase diagram. The behavior of the electric charge and strangeness chemical potentials in the presence of rotation is also analyzed, providing the first systematic study of their rotational dependence within the HRG framework. Furthermore, we examine the impact of rotation on experimentally relevant observables, including hadron yield ratios and susceptibility ratios of conserved charges. Our results show that while particle yield ratios exhibit noticeable sensitivity to rotation, the conventional cumulant ratios remain comparatively less affected. This indicates that hadronic yield ratios may provide a more suitable observable for estimating the magnitude of rotational effects generated in heavy-ion collisions.

\end{abstract}
\date{\today}
\maketitle
\section{Introduction}
One of the primary objectives of various heavy-ion collision programs, such as those conducted at the Relativistic Heavy Ion Collider (RHIC), is to understand the properties of quantum chromodynamics (QCD) matter at extreme conditions and hence, to map the QCD phase diagram in the temperature ($T$) and baryochemical potential ($\mu_B$) plane. External fields, such as strong magnetic fields produced in non-central heavy-ion collisions, can significantly influence the properties of the QCD medium~\cite{Huang:2015oca, Bali:2011qj, Tawfik:2016lih}. Besides giving rise to intriguing phenomena such as the chiral magnetic effect~\cite{Kharzeev:2007jp, Fukushima:2008xe}, the magnetic field can lead to a substantial change in the quark-hadron transition temperature~\cite{Bali:2011qj}. Furthermore, mechanisms such as magnetic catalysis~\cite{Kharzeev:2013jha} and inverse magnetic catalysis~\cite{Bruckmann:2013oba} can substantially modify the structure of the QCD phase diagram. Recent studies have also suggested the possibility of a critical point in the temperature–magnetic field plane~\cite{Endrodi:2015oba, DElia:2021yvk}. In recent years, another important aspect that has attracted considerable attention is the possibility of generating a large amount of vorticity (rotation) in peripheral heavy-ion collisions. The observation of finite hyperon polarization at the STAR experiment~\cite{STAR:2017ckg} indicates that the produced medium can possess vorticity ($\omega$) of the order of $10^{21}$ s$^{-1}$, making it the most vortical fluid ever observed in nature. Such large rotational effects can significantly influence the properties of QCD matter and introduce an additional dimension to the study of the QCD phase diagram. In this context, several theoretical studies have explored the role of rotation in strongly interacting matter. For instance, Ref.~\cite{Jiang:2016wvv} predicts the possible existence of a critical point associated with the chiral phase transition in the $T$–$\omega$ plane. Similarly, Ref.~\cite{Pradhan:2023rvf} reports the emergence of a liquid–gas critical point in hadronic matter in the $T$–$\omega$ plane at $\mu_B = 0$. Moreover, the impact of rotation on various properties of hadronic matter has been investigated, including its thermodynamic behavior~\cite{Pradhan:2023rvf, Fujimoto:2021xix, Mukherjee:2023qvq}, higher-order fluctuations of conserved charges~\cite{Mukherjee:2023ijv, Sahoo:2025fif}, and transport properties of the medium~\cite{Padhan:2024edf,Padhan:2025qhz,Dwibedi:2025boz}.

In the past, statistical hadronization models, such as the hadron resonance gas (HRG) model, based on thermal hadron distributions, were found to be quite successful in describing hadron yields from high-energy collision experiments~\cite{Braun-Munzinger_inBook}. Within this framework, the system at chemical freeze-out can be characterized by two free thermodynamic parameters, the chemical freeze-out temperature ($T_{ch}$) and the corresponding baryochemical potential (${\mu_B}_{ch}$). Over a wide range of collision energies, this model is used to extract the freeze-out curve in the $T-\mu_B$ plane~\cite{Cleymans:2005xv}. 
In addition, the thermodynamic susceptibilities in these frameworks can be related to conserved-charge fluctuation measurements, expressed in terms of cumulants, in experiments. It has been observed that the particle susceptibility of the lowest order, calculated in the HRG model, is in fair agreement with the experimental data and also allows us to extract the freeze-out parameters directly from the experiments~\cite{Bazavov:2012vg, Borsanyi:2013hza}. Moreover, in Ref.~\cite{Alba:2015iva}, the authors have compared the two methods for extracting the freeze-out parameters, namely particle yields and fluctuations of conserved charge, within the framework of the HRG model. They show that for certain hadrons, particle-multiplicity fluctuations are more reliable for estimating the freeze-out temperature, whereas for others, both methods yield comparable results. In addition, several phenomenological conditions have been proposed in the literature to determine the chemical freeze-out parameters within the HRG framework. Certain characteristics of the thermal medium, such as the average energy per particle at chemical freeze-out, exhibit similar features across different collision energies. These common features suggest the existence of a unified set of freeze-out conditions in heavy-ion collisions. In Ref.~\cite{Cleymans:1998fq, Cleymans:1999st}, it has been shown that the freeze-out parameters obtained in different collision energies correspond to a unique constant value of average energy per hadron. Thus came the condition of fixed energy per hadron, $ E/ N \sim$ 1.08 GeV as the criterion for the chemical freeze-out parameter in heavy-ion collisions. There have been several other proposals for this, such as a constant value of the temperature-scaled entropy density ($s/T^{3}\sim7$)~\cite{Cleymans:2004hj, Tawfik:2004vv}, a fixed total baryon and anti-baryon density~\cite{Braun-Munzinger:2001uhy}, etc. A detailed study of these conditions and their comparison is presented in Ref.~\cite{Cleymans:2005xv}. It was shown that the constant energy per hadron criterion, $ E/ N$, provides a more reliable description of chemical freeze-out over a wide range of collision energies.

In recent years, the possible impact of external fields on the chemical freeze-out conditions has also been studied. In particular, strong magnetic fields generated in non-central heavy-ion collisions can modify the thermodynamic properties of strongly interacting matter~\cite{Tawfik:2016lih, Sahoo:2023vkw} and influence particle yields and fluctuations~\cite{Vovchenko:2024wbg, Marczenko:2024kko}, thereby directly affecting the freeze-out conditions. In the HRG framework, the authors in Ref.~\cite{Fukushima:2016vix} have used the criterion of $E / N $ in the range of 0.9 to 1.0 to obtain the chemical freeze-out line in the presence and absence of a magnetic field in the system. It has been observed that the magnetic field shifts the freeze-out line towards lower temperatures, demonstrating the characteristic of inverse magnetic catalysis. They also predicted that in the presence of a strong magnetic field, the electric charge fluctuations are significantly enhanced at high baryon density. Further, the lattice QCD (lQCD) computations of conserved charge fluctuations and correlation~\cite{Ding:2023bft, Ding:2025jfz} have demonstrated that the net baryon and electric charge correlation ($\chi^{BQ}_{11}$) is indeed strongly sensitive to the magnetic field and can act as a magnetometer of QCD. 

Motivated by the significant impact of the magnetic field on the freeze-out conditions, we investigate in this work whether rotation, which is analogous to a magnetic field in many respects, can also affect the chemical freeze-out parameters in heavy-ion collisions. Analogously to the magnetic field, the rotation of the system alters the effective chemical potentials of the particles, which can in turn influence the thermodynamic conditions at freeze-out. The authors in Ref.~\cite{Mukherjee:2023qvq} have attempted to study the combined effect of a magnetic field and rotation on the freeze-out temperature. They have estimated the freeze-out curve based on the rapid increase in thermodynamic quantities such as entropy density and from the shift of the dip in the squared speed of sound as a function of temperature. Their results indicate a shift of the freeze-out curve toward lower temperatures in the presence of both rotation and a magnetic field. However, a systematic investigation of the rotational effect on hadronic abundances, fluctuation observables, and the determination of freeze-out parameters is still lacking in the literature. In this work, we consider a rotating hadronic medium to examine the role of rotation in determining freeze-out conditions in heavy-ion collisions. In particular, for the first time, we determine the freeze-out curve using the commonly employed constant energy per hadron condition in the presence of rotation. We also employ the normalized entropy density criterion to compare the freeze-out parameters obtained from different conditions. Furthermore, we investigate the effect of rotation on hadronic abundances by evaluating particle yield ratios as a function of rotation in a rotating HRG medium. In addition, we calculate higher-order fluctuations of conserved charges and their ratios to examine their behavior under rotational effects. This naturally allows us to compare the sensitivity of hadron yield ratios and fluctuation observables to rotation, which may provide useful insights for extracting the magnitude of vorticity produced in heavy-ion collisions. It is noteworthy to mention that to obtain the freeze-out parameters ($T_{ch},{\mu_{B}}_{ch}$), it is necessary to fix the electric charge ($\mu_Q$) and strangeness ($\mu_S$) chemical potential that characterizes the thermal medium created in high-energy collisions. These chemical potentials can be determined by imposing constraints corresponding to the conserved quantum numbers of the colliding nuclei, namely strangeness neutrality and a fixed ratio of electric charge to baryon number of the system~\cite{Bazavov:2012vg,Lysenko:2024hqp}. Since magnetic fields modify hadronic abundances, this should affect these freeze-out conditions, which then acquire an explicit magnetic-field dependence, as demonstrated in Ref.~\cite{Fukushima:2016vix}. By analogy, a similar modification is therefore expected in the presence of rotation. In this work, we evaluate these effects within a rotating medium to gain a deeper understanding of how rotation influences the freeze-out criteria and look for a better observable to extract the magnitude of rotation produced in heavy-ion collision experiments.

The structure of the present article is as follows. In Sec.~\ref{sec_formulation}, we present the formulation of a rotating HRG model and discuss its thermodynamic properties along with the relations required for calculating higher-order conserved-charge fluctuations. In Sec.~\ref{sec_results}, we discuss the freeze-out criteria and analyze the impact of rotation on the freeze-out parameters with reference to the obtained results. In the same section, we also examine hadron yield ratios and fluctuation observables in the presence of rotation. Finally, we summarize our main findings in Sec.~\ref{sce_summary}.

\section{Formalism}
\label{sec_formulation}
The HRG model consists of a system of experimentally established hadrons and resonances. This model is quite successful in explaining the experimental results on particle yields~\cite{ALICE:2022wpn}. It also explains the lQCD results on thermodynamics at low temperature~\cite{Bellwied:2013cta, HotQCD:2012fhj, Bellwied:2017ttj}. In the Grand Canonical Ensemble (GCE) framework, the total pressure of the given hadronic system can be calculated by summing the pressure due to each hadronic species as~\cite{Pradhan:2023rvf, Andronic:2012ut}
\begin{equation}
\label{eq_normalP}
    P^{id}_i(T,\mu_i) = \pm \frac{Tg_i}{2\pi^2} \int_{0}^{\infty} p^2 dp\ \ln\{1\pm \exp[-(E_i-\mu_i)/T]\},
\end{equation}
where the quantities $g_i$ and $E_i = \sqrt{p^2 + m_i^2}$ represent the degeneracy factor and the energy of the $i$th hadron, respectively. The $\pm$ sign corresponds to fermions and bosons. The chemical potential associated with the $i$th hadron is denoted by $\mu_i$ and is expressed as
\begin{equation}
\label{eq_mu}
    \mu_i = B_i\mu_B + S_i\mu_S +Q_i\mu_Q,
\end{equation}
where $\mu_B$, $\mu_S$, and $\mu_Q$ denote the baryon, strangeness, and electric charge chemical potentials, respectively. The quantities $B_i$, $S_i$, and $Q_i$ correspond to the baryon number, strangeness, and electric charge of the $i$th hadron.

When the system rotates with angular velocity $\omega$ along a direction, say, the $z$ axis, the single–particle energy spectrum is modified due to the coupling between rotation and particle angular momentum. In this case, the dispersion relation receives a rotational contribution and can be written as $\varepsilon_{l} = E-(l+s)\omega$, where $s$ denotes the $z$ component of the spin and $l$ represents the orbital angular momentum quantum number along the $z$ axis. Under such conditions, the pressure due to $i$th hadronic species in a rotating medium takes the form~\cite{Fujimoto:2021xix, Mukherjee:2023qvq, Pradhan:2025pol}
\begin{align}
\label{eq_pressure}
    P_i &= \pm\frac{T}{8\pi^2} \sum_{\ell=-\infty}^\infty \int dk_r^2 \int  dk_z\; \sum_{\nu = \ell}^{\ell + 2S_i}  J_\nu^2(k_r r) \notag\\
    &\qquad\qquad \times \log\left\{1\pm\exp[-(\varepsilon_{\ell,i} - \mu_i) / T]\right\},
\end{align}

where the single–particle energy in the rotating frame is expressed as $\varepsilon_{l,i}$ = $\sqrt{k_r^2 +k_z^2 +m_i^2}-(l+s)\omega$. Here, $r$ denotes the radial distance from the rotation axis, while $J_\nu$ represents the Bessel function of the first kind. The transverse and longitudinal momentum components are given by $k_r$ and $k_z$, respectively. To ensure causality, a boundary condition must be imposed at $r=R$ such that $R\omega\leq1$ ($c=1$ in natural units). Throughout this work, we fix $R$ = 30 GeV$^{-1}$ ($\approx$ 6 fm). The presence of this boundary condition leads to the quantization of the radial momentum as $k_r=\xi_{l,i}/R$~\cite{Fujimoto:2021xix}, where $\xi_{l,i}$ corresponds to the $i$th zero of the Bessel function satisfying $J_l(\xi_{l,i})=0$. As a consequence, momentum is discretized, mainly affecting the low-momentum region. Therefore, the lower limit of the integration over $k_r$ is no longer zero but effectively acquires a finite value given by $\xi_{l,i}/R=\xi_{l,i}\omega$.
The Bessel function $J_\nu(k_r r)$ incorporates an explicit radial dependence into the formalism of the rotating system. Following Refs.~\cite{Fujimoto:2021xix,Pradhan:2025pol}, we have evaluated all the thermodynamic quantities at a fixed radial position. This prescription avoids ambiguities arising from spatial variations and enables clearer interpretation of rotational effects. 
All numerical calculations are performed at the boundary of the system, $r = R$ = 30 GeV$^{-1}$, in order to probe the maximal effects of rotation.

From Eq.~(\ref{eq_pressure}), one can obtain the total number density, energy density, and entropy density of the hadronic system as
\begin{align}
    \label{eq_numden}
    n &= \bigg(\frac{\partial P}{\partial \mu}\bigg)_{T,\omega}\nonumber \\
    &= \frac{1}{8\pi^2} \sum_{\ell=-\infty}^\infty \int dk_r^2 \int  dk_z\; \sum_{\nu = \ell}^{\ell + 2S_i}  J_\nu^2(k_r r) \notag\\
    &\qquad\qquad \times \frac{1}{\exp[(\varepsilon_{\ell,i} - \mu_i) / T]\pm 1}.
\end{align}

\begin{align}
    \label{eq_energy_den}
    \varepsilon &= -\frac{1}{T}\bigg(\frac{\partial P}{\partial \frac{1}{T}}\bigg)_{\frac{\mu}{T},\frac{\omega}{T}}\nonumber \\
    &= \frac{1}{8\pi^2} \sum_{\ell=-\infty}^\infty \int dk_r^2 \int  dk_z\; \sum_{\nu = \ell}^{\ell + 2S_i}  J_\nu^2(k_r r) \notag\\
    &\qquad\qquad \times \frac{\varepsilon_{\ell,i}}{\exp[(\varepsilon_{\ell,i} - \mu_i) / T]\pm 1}.
\end{align}

\begin{align}
s = \left(\frac{\partial P}{\partial T}\right)_{\mu,\omega} =& \pm \frac{1}{8\pi^2} \sum_{\ell=-\infty}^{\infty} \int dk_r^2 \int dk_z 
     \sum_{\nu=\ell}^{\ell+2S_i} J_\nu^2(k_r r)  \notag\\
  &\times \Bigg[\log\left\{1\pm\exp\!\left(-\frac{\varepsilon_{\ell,i}-\mu_i}{T}\right)\right\} \notag\\
     &\quad \quad \quad \pm \frac{\varepsilon_{\ell,i}-\mu_i}{T}
     \frac{1}{\exp\!\left(\frac{\varepsilon_{\ell,i}-\mu_i}{T}\right)\pm 1}\Bigg].
\end{align}
We can use these equations for the freeze-out criterion of constant energy per particle, $ E / N $ $\simeq$ $\varepsilon /n $ = 1.08 GeV, and the constant entropy density $s/T^3$ = 7. The fluctuations and correlations of conserved charges in a thermally and chemically equilibrated HRG can be characterized through generalized susceptibilities. These quantities are obtained by taking derivatives of the scaled pressure in Eq. (\ref{eq_pressure}) with respect to the corresponding chemical potentials. The susceptibilities are defined as~\cite{Ding:2025nyh}

\begin{equation}
\label{equation_susceptibility}
\chi^{BQS}_{klm} = \frac{\partial^{k+l+m}[P(\mu_{B}, \mu_{Q}, \mu_{S})/T^{4}]}{\partial \mu_{B}^{k} ~\partial \mu_{Q}^{l} ~\partial \mu_{S}^{m}}\bigg |_{\mu = 0},
\end{equation}

where $k$, $l$, and $m$ denote integers specifying the order of the derivatives with respect to the baryon, electric charge, and strangeness chemical potentials, respectively. Using Eq.~(\ref{equation_susceptibility}), we can calculate fluctuations of conserved charges such as baryon, electric charge, and strangeness, along with the correlation among different charges in a rotating hadron gas system.

\begin{figure*}[htp!]
\begin{center}
\includegraphics[scale = 0.5]{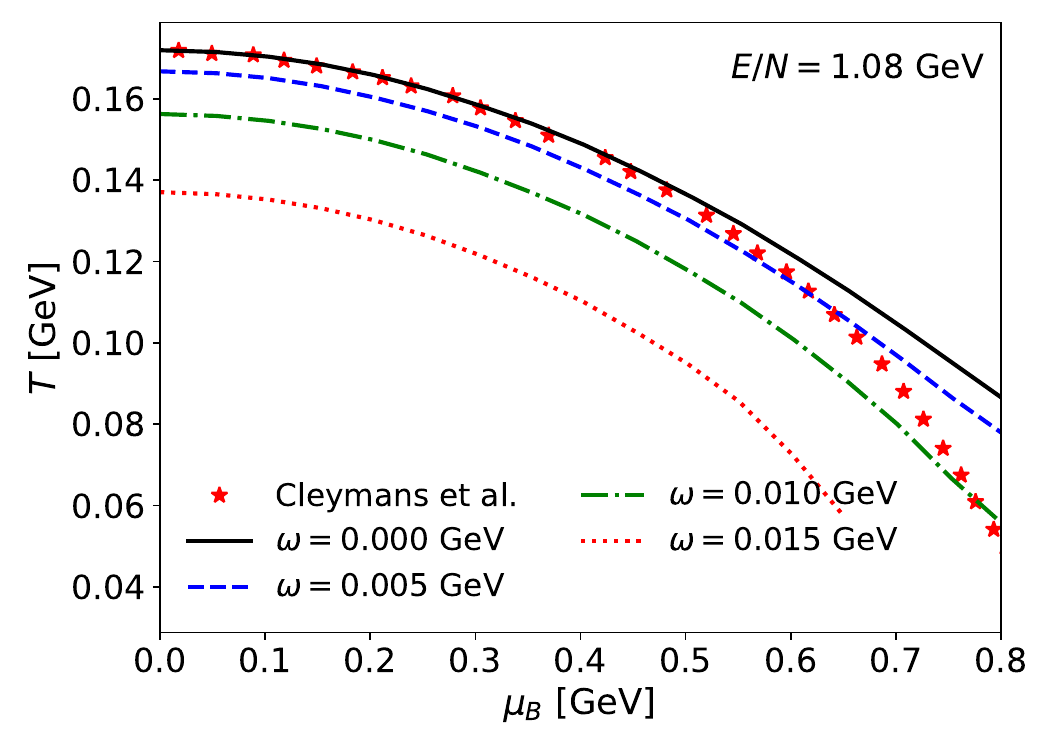}
\includegraphics[scale = 0.5]{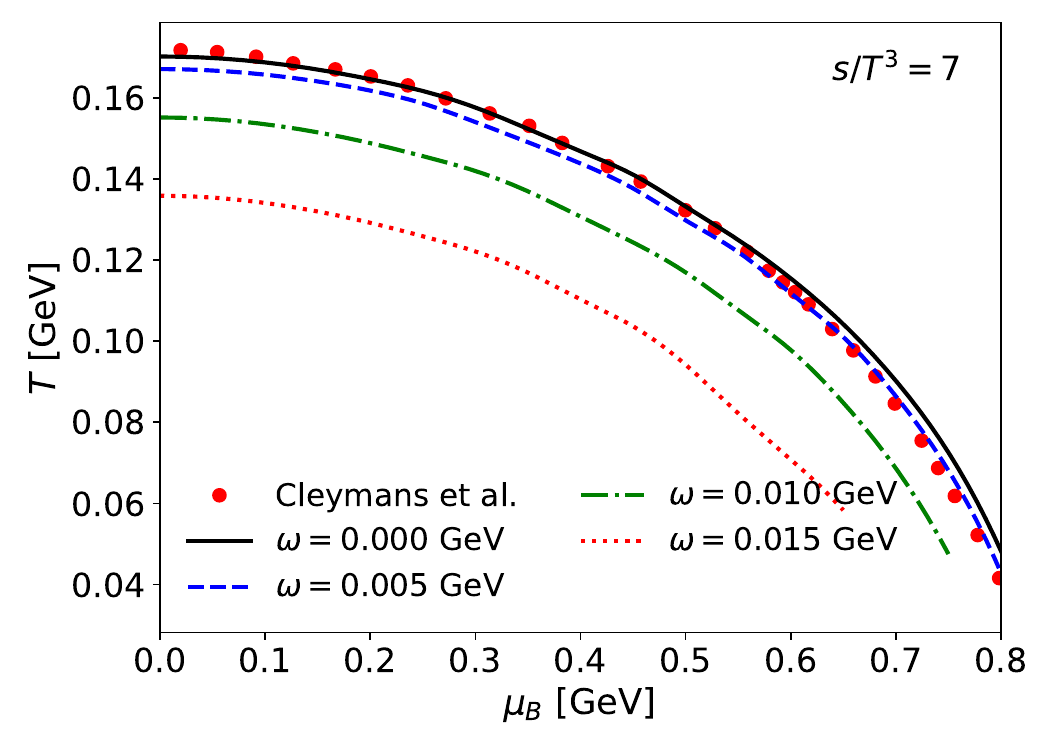}
\caption{(Colour Online) The freeze-out curve in the $T-\mu_B$ plane is obtained using the conditions of constant average energy per particle, $ \varepsilon / n  = 1.08~\mathrm{GeV}$ (left), and constant entropy density, $s/T^3 = 7$ (right). The results at $\omega = 0$ are compared with those reported by Cleymans et al.~\cite{Cleymans:2005xv}, and are further extended to finite rotation with $\omega = 0.005$, $0.01$, and $0.015~\mathrm{GeV}$.
}
\label{fig1}
\end{center}
\end{figure*}

\section{Results and Discussion}
\label{sec_results}

Statistical hadronization models have long been successful in describing particle production in heavy-ion collisions, assuming a chemical freeze-out surface at which inelastic particle production ceases and the hadronic composition becomes fixed. 
The final particle yields are further shaped by the decay of heavier resonances. For a given collision energy, the system is characterized by specific values of $T$ and $\mu_B$. It has been observed that certain thermodynamic variables at chemical freeze-out exhibit near-universal behaviour, suggesting the existence of common freeze-out conditions~\cite{Cleymans:2005xv}. 
The existence of such criteria enables a systematic study of particle production and its energy dependence in heavy-ion collisions. In this work, we employ the HRG model to revisit these conditions and to study the possible impact of rotation on them. We incorporate all hadrons and resonances into the model with masses up to 2.6~GeV~\cite{ParticleDataGroup:2016lqr}.

In this context, empirical freeze-out criteria based on combinations of thermodynamic quantities have been particularly successful in characterizing chemical freeze-out. One such widely used condition is the constant average energy per particle, $\varepsilon/n \approx 1.08~\text{GeV}$, as it successfully reproduces hadron yield over a wide range of collision energies. It signifies the stage at which the system no longer possesses sufficient energy to sustain inelastic interactions, thereby fixing particle abundances~\cite{Cleymans:1998fq,Cleymans:2005xv}. 
An alternative and widely used criterion is the normalized entropy density, $s/T^3 \approx 7$, which remains approximately constant across different collision energies. This behavior suggests that the effective number of thermodynamic degrees of freedom at freeze-out is nearly universal, providing a consistent and independent condition for determining the freeze-out parameters within the HRG framework~\cite{Cleymans:2005xv}. To construct this freeze-out curve, the electric charge and strangeness chemical potentials ($\mu_Q$, $\mu_S$) are introduced by imposing the conditions of charge conservation and strangeness neutrality. To ensure overall charge conservation, the ratio of net electric charge to net baryon number, $n_Q/n_B$, is fixed by the initial isospin asymmetry of the colliding system, typically taken as $Z/A$ of the nuclei, which is approximately 0.4 for heavy nuclei like Pb and Au. Hence the quantities $\mu_Q$ and $\mu_S$ are fixed by the constraints of
\begin{align}
\frac{n_Q(T,\mu_B,\mu_Q,\mu_S, \omega)}{n_B(T,\mu_B,\mu_Q,\mu_S, \omega)} &= 0.4, \label{eq:freeze2} \\
n_S(T,\mu_B,\mu_Q,\mu_S, \omega) &= 0. \label{eq:freeze3}
\end{align}

In Fig.~\ref{fig1}, we portray the freeze-out curve using $\varepsilon/n = 1.08~\mathrm{GeV}$ (left panel) and  $s/T^3 = 7$ (right panel) to study the effect of rotation on freeze-out dynamics. The black solid line represents the freeze-out curve at $\omega = 0$ and is compared with the results for the non-rotating case from Ref.~\cite{Cleymans:2005xv}, shown by the red markers. Our result in the non-rotating limit is also consistent with the findings of Lysenko et al.~\cite{Lysenko:2024hqp}, corresponding to $\varepsilon/n = 0.951~\text{GeV}$. We then proceed to study the effects of rotation on the $T\text{--}\mu_B$ curve within the framework of the rotating HRG model, as described in Sec.~\ref{sec_formulation}. 
We consider different values of the rotational strength $\omega = 0.005$, $0.01$, and $0.015$~GeV with the corresponding curves represented by the blue dashed, green dash-dotted, and red dotted lines, respectively. 
Earlier in Ref.~\cite{Fukushima:2016vix}, the authors studied the influence of the magnetic field on the freeze-out curve using the constant average energy per particle criterion. Their results show a downward shift of the freeze-out curve, indicating inverse magnetic catalysis behavior. 
In the present study, we find that rotation leads to a similar modification of the freeze-out curve. 
In a rotating medium, the single particle energy is modified, which in turn affects the effective chemical potential, as seen in Eq.(\ref{eq_pressure}). 
For $\omega = 0.005$~GeV, the freeze-out temperature decreases by approximately $5$~MeV and $3$~MeV for both the $\varepsilon/n$ and $s/T^3$ conditions, respectively. 
This freeze-out temperature is further reduced with increasing rotational magnitude. It is evident that rotation contributes to the effective chemical potential and lowers the freeze-out temperature. This may have important implications for future experimental analyses of particle yields and fluctuations in heavy-ion collisions, which could provide insights into rotational effects and offer a possible avenue to test these theoretical predictions. In fact, it shows that the effect of rotation cannot be ignored when estimating freeze-out parameters from experiments.

\begin{figure}[htp!]
\begin{center}
\includegraphics[scale = 0.65]{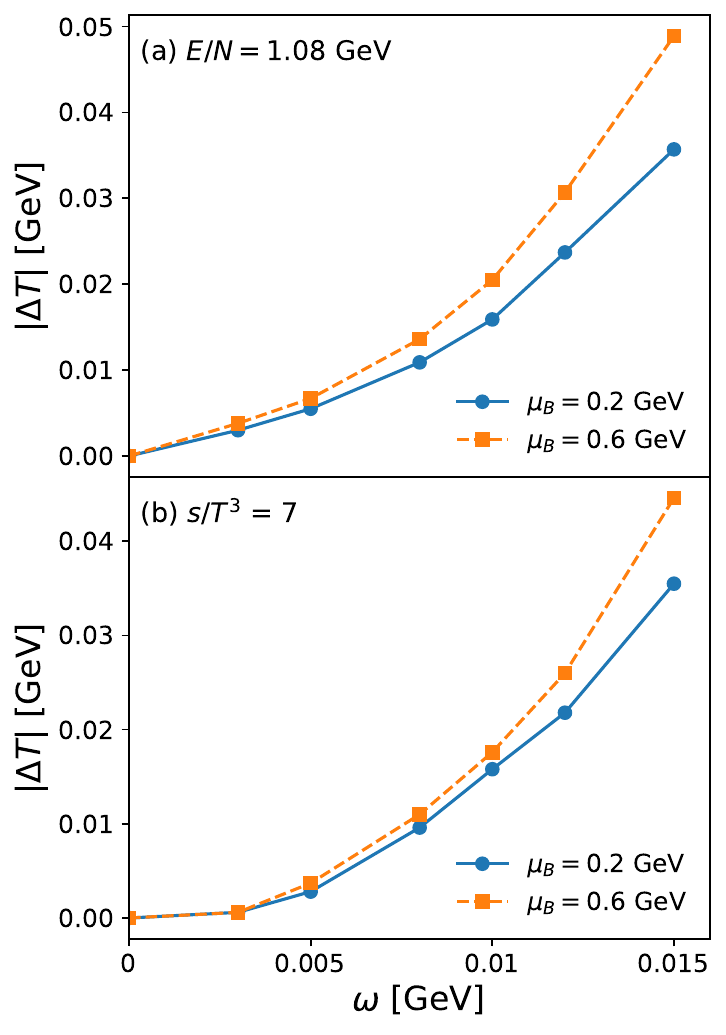}
\caption{(Colour Online) 
The shift in freeze-out temperature ($\Delta T$) as a function of $\omega$ for $\mu_B = 0.2$ and $0.6$~GeV.}
\label{figure2}
\end{center}
\end{figure}

To study the impact of rotation on the shift in freeze-out temperature ($\Delta T$), we estimate its dependence on angular velocity. In Fig.~\ref{figure2}, we present the variation of $\Delta T$ as a function of rotational angular velocity for two representative values of baryon chemical potential, $\mu_B = 0.2$ and $0.6$~GeV. The upper panel corresponds to the shift with freeze-out condition based on the $\varepsilon/n$ criterion, while the lower panel presents the results using the $s/T^3$ criterion. For $\omega = 5$ MeV, the temperature shift is small, of the order of $\sim 5$ MeV. As $\omega$ increases, the shift becomes more pronounced, reaching approximately $40$–$50$ MeV at $\omega = 15$ MeV. This indicates that the effect of rotation on the freeze-out temperature grows with increasing $\omega$ and exhibits a non-linear dependence, as illustrated in the figure. Here, the solid and dashed lines correspond to $\mu_B = 0.2$ and $0.6$~GeV, respectively. The impact of rotation on the temperature shift for different values of the baryon chemical potential is also clearly visible. At low rotational strength, the shift in temperature remains relatively insensitive to the baryon chemical potential, $\mu_B$. However, as the rotational strength increases, the magnitude of the temperature shift becomes larger and exhibits a clear dependence on $\mu_B$. In particular, higher values of $\mu_B$ lead to increasingly distinct deviations in the temperature shift at larger rotation, indicating an enhanced interplay between the effects of rotation and baryon density.

\begin{figure*}[htp!]
\begin{center}
\includegraphics[scale = 0.55]{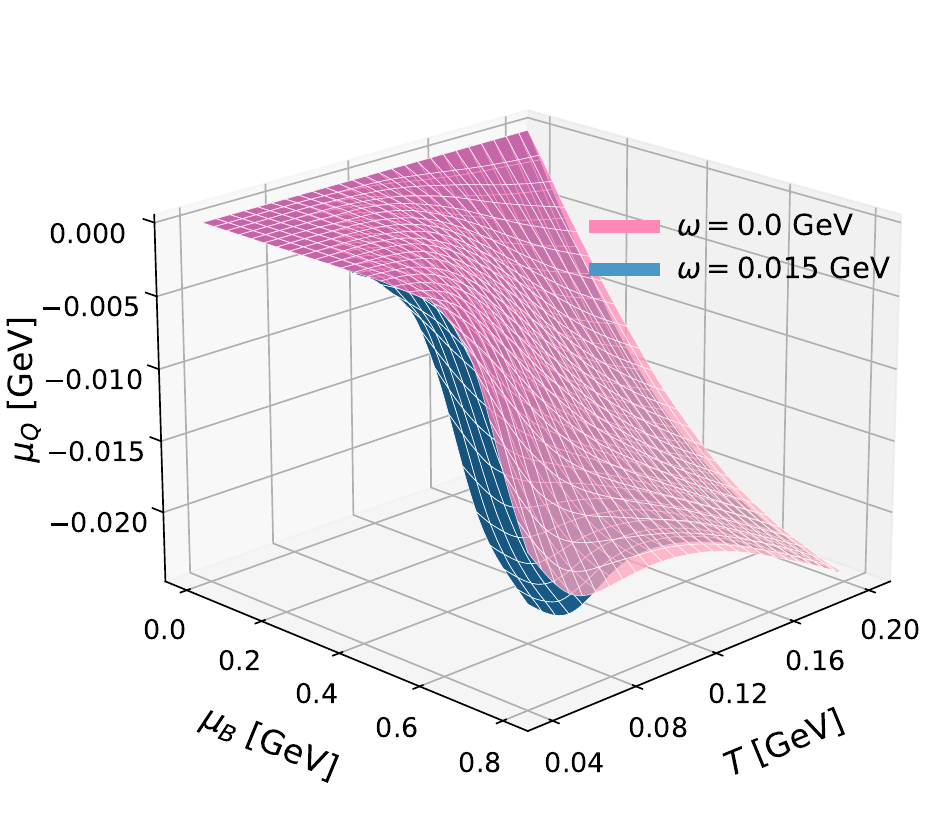}
\includegraphics[scale = 0.55]{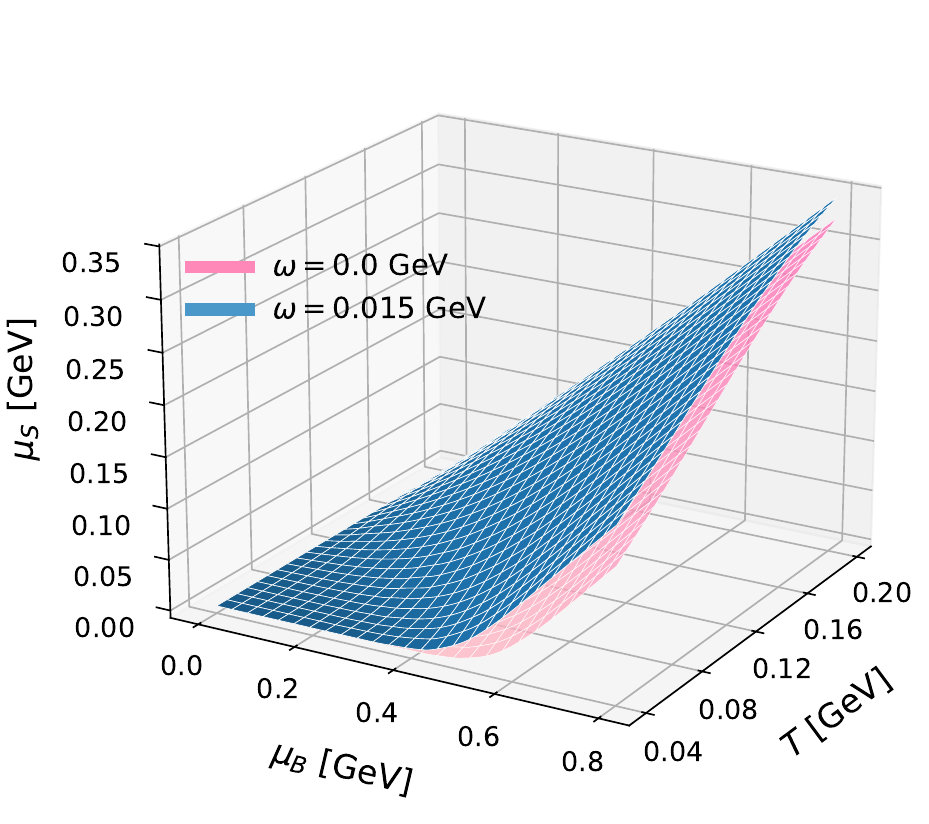}
\caption{(Colour Online) The electric charge and strangeness chemical potential as a function of T and $\mu_B$ for $\omega$ = 0 and 0.015 GeV.}
\label{figure3}
\end{center}
\end{figure*}

As discussed earlier, fixing $\mu_Q$ and $\mu_S$ is essential to ensure charge conservation and to reflect realistic experimental conditions. Moreover, in the presence of rotation, these chemical potentials are expected to be modified. In obtaining the freeze-out curves shown in Fig.~\ref{fig1}, these parameters are determined consistently by incorporating finite rotational effects. However, to gain further insight, we now explicitly investigate the impact of rotation on these parameters, specifically the charge and strangeness chemical potentials at freeze-out. In Fig.~\ref{figure3}, we show the variation of $\mu_Q$ and $\mu_S$ as a function of $T$ and $\mu_B$ for two different values of angular velocity. These parameters are obtained by solving the constraint equations given in Eqs.~(\ref{eq:freeze2}) and (\ref{eq:freeze3}), which enforce a fixed charge-to-baryon ratio and strangeness neutrality. For $\omega = 0$ curve, the charge chemical potential $\mu_Q$ decreases with increasing $T$ and $\mu_B$, following the approximate relation $\mu_Q \sim -\mu_B/30$. In contrast, the strangeness chemical potential $\mu_S$ increases with $T$ and $\mu_B$, obeying the relation $\mu_S \sim \mu_B/3$ as in Ref.~\cite{Lysenko:2024hqp}. One can say that this behavior arises from the requirement to maintain the conserved charge conditions as the baryon density increases. Upon introducing rotation into the system, an additional contribution arises in the form of an effective chemical potential associated with the rotational motion. For a rotational strength of $\omega = 0.015$~GeV, the magnitude of $\mu_Q$ is further reduced, while $\mu_S$ shows a more pronounced increase compared to the non-rotating case. This enhancement can be attributed to the rotational effects, which modify the particle phase space and thermodynamic distributions, thereby requiring larger chemical potentials to satisfy the same conservation constraints. This behavior of $\mu_Q$ and $\mu_S$ with rotation is found to be analogous to that of the magnetic field as shown in Ref.~\cite{Fukushima:2016vix}.

At fixed values of $(T, \mu_B)$, the electric charge and strangeness chemical potentials $(\mu_Q, \mu_S)$, consistent with the constraints in Eqs.~(\ref{eq:freeze2}) and (\ref{eq:freeze3}), can be determined within QCD~\cite{Bazavov:2012vg}. This is achieved by performing a Taylor expansion of the conserved charge densities in terms of the three chemical potentials. The corresponding expansion coefficients are evaluated using lQCD, which involves the numerical computation of generalized susceptibilities in Eq. (\ref{equation_susceptibility}). The chemical potentials can be expressed to the next-to-leading order (NLO) in $\mu_B$ as~\cite{Bazavov:2012vg, Ding:2025nyh}
\begin{equation}
\mu_Q = q_1 \mu_B + q_3 \mu_B^3, \qquad
\mu_S = s_1 \mu_B + s_3 \mu_B^3.
\end{equation}
where the leading order (LO) coefficients are expressed in terms of susceptibilities as \cite{Bazavov:2012vg, Ding:2025nyh} 
\begin{align}
q_1 &= 
\frac{
t \left( \chi_{2}^{B}\chi_{2}^{S} - \chi_{11}^{BS}\chi_{11}^{BS} \right)
-
\left( \chi_{11}^{BQ}\chi_{2}^{S} -  \chi_{11}^{BS}\chi_{11}^{QS} \right)
}{
\left( \chi_{2}^{Q}\chi_{2}^{S} - \chi_{11}^{QS}\chi_{11}^{QS} \right)
-
t \left( \chi_{11}^{BQ}\chi_{2}^{S} - \chi_{11}^{BS}\chi_{11}^{QS} \right)},
\label{q1}\\[10pt]
s_1 &= - \frac{\chi_{11}^{BS}}{\chi_{2}^{S}} - \frac{\chi_{11}^{QS}}{\chi_{2}^{S}} q_1.
\label{s1}
\end{align}

\begin{figure*}[htp!]
\begin{center}
\includegraphics[scale = 0.5]{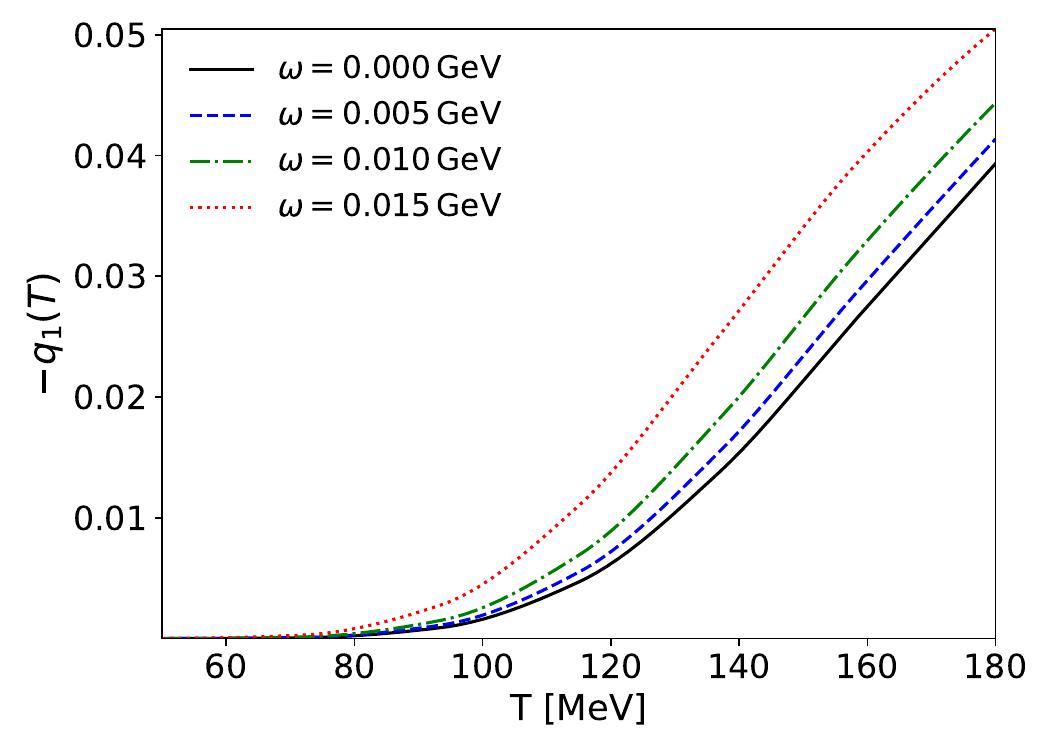}
\includegraphics[scale = 0.5]{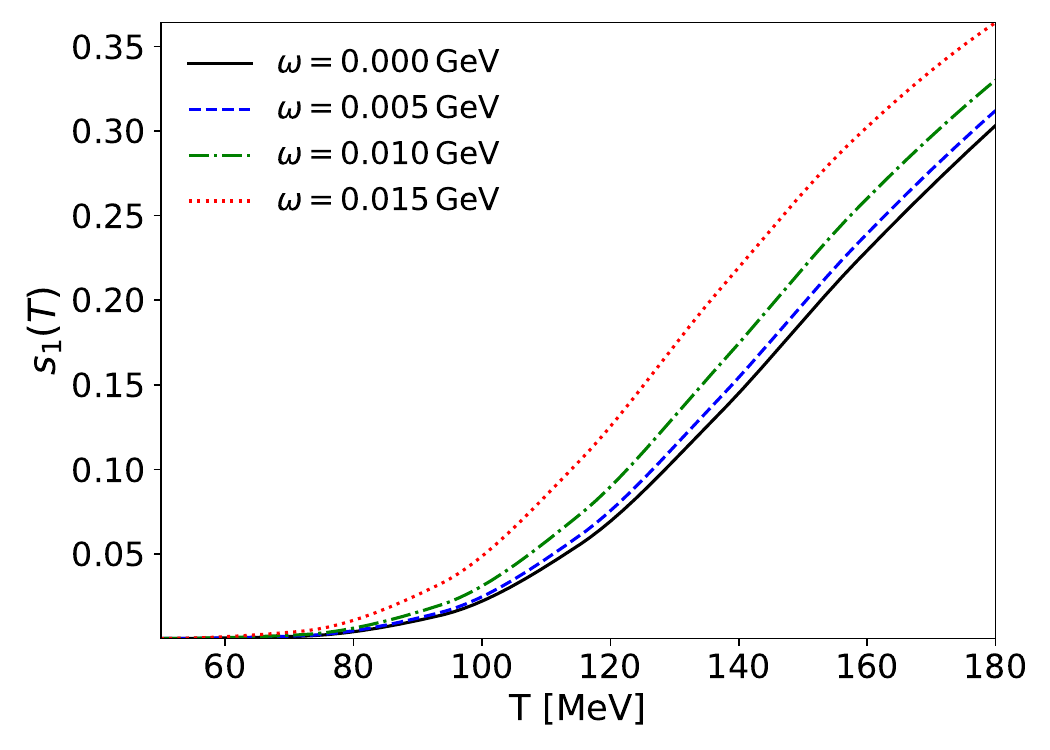}
\caption{(Colour Online) The leading-order expansion coefficients of the negative of electric charge (left) and strangeness chemical potentials (right) as a function of temperature at $\mu_B$ = 0 for different magnitudes of rotation.}
\label{figure4}
\end{center}
\end{figure*}

\begin{figure*}[ht!]
\begin{center}
\includegraphics[scale = 0.5]{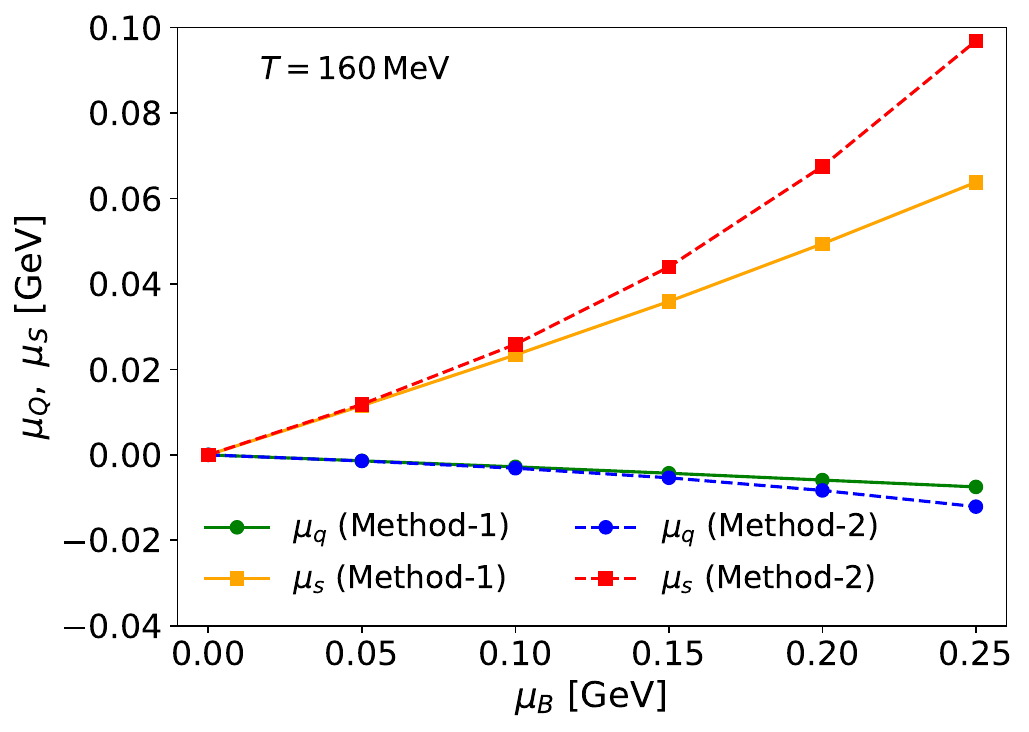}
\includegraphics[scale = 0.5]{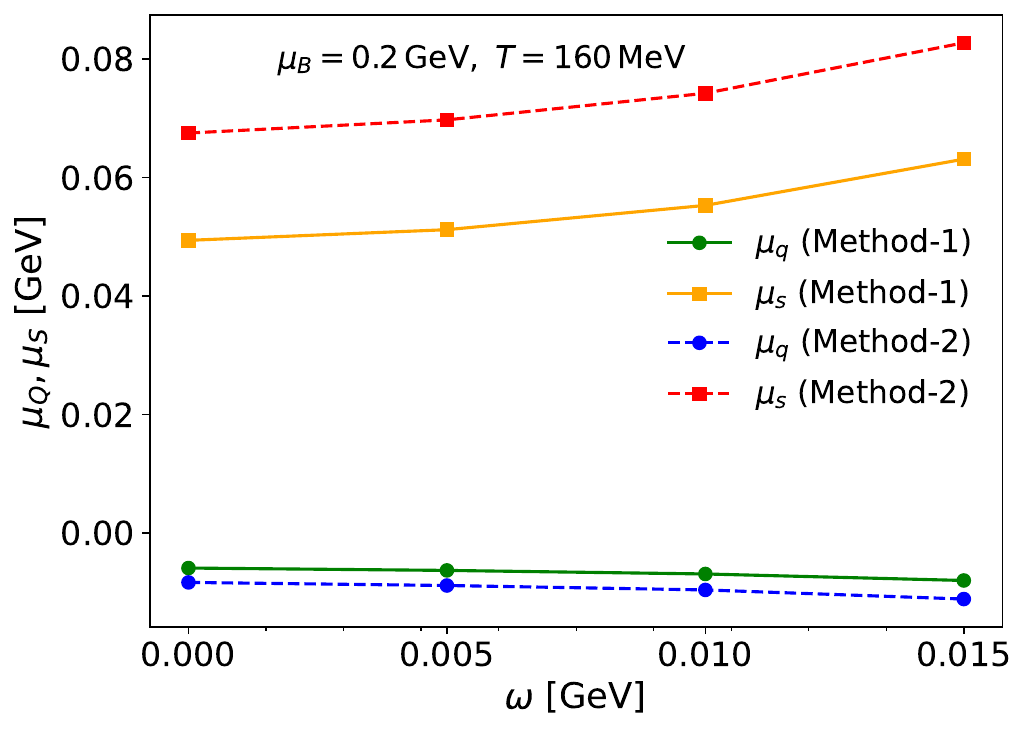}
\caption{(Colour Online) The chemical potentials $\mu_Q$ and $\mu_S$ obtained using two different approaches are compared as a function of $\mu_B$ at $\omega$ = 0 (left) and as a function of $\omega$ at $\mu_B$ = 0.2 GeV and T = 160 MeV (right).}
\label{figure5}
\end{center}
\end{figure*}

where $t=n_Q/n_B$ = 0.4 as mentioned in Eq.(\ref{eq:freeze2}). Within the HRG framework, the charge and strangeness chemical potentials can be estimated directly by using Eq.~(\ref{eq:freeze2}) and (\ref{eq:freeze3}), which we refer to as method-1. Alternatively, these quantities can be determined by evaluating the leading-order coefficients $q_1$ and $s_1$, as done in lQCD calculations~\cite{Bazavov:2012vg}, using thermodynamic susceptibilities. This approach is referred to as method-2 here. We have calculated this $q_1$ and $s_1$ using Eq.~(\ref{q1}) and (\ref{s1}), where susceptibilities are calculated from Eq.~(\ref{equation_susceptibility}) in the limit of vanishing, baryon, electric charge, and strangeness chemical potentials. 
Previous lQCD calculations~\cite{Bazavov:2012vg} have shown that the NLO expansion reproduces the full HRG results for $(\mu_Q, \mu_S)$ with an accuracy better than $1\%$ for $\mu_B/T \leq 1.3$ and the approximation is adequate to describe the overall behavior of the results. In Ref.~\cite{Ding:2025nyh}, the authors have reported the effect of the magnetic field on the LO coefficient of the QCD equation of state. As a first step in this study, we investigate the impact of rotation on the LO coefficients, $q_1$ and $s_1$. Fig.~\ref{figure4} shows their temperature dependence for different values of the angular velocity. The black solid line corresponds to the non-rotating limit, whereas the blue dashed, green dashdot, and red dotted lines correspond to $\omega$ = 0.005, 0.01, 0.015 GeV, respectively. We observe from the left panel of Fig.~\ref{figure4} that $q_1 \equiv q_1(T,\,\omega) = (\mu_Q/\mu_B)_{\mathrm{LO}}$ remains negative over the entire $T-\omega$ parameter space. It is therefore presented with an explicit negative sign. The inclusion of rotation further enhances the magnitude of $q_1$, making it more negative as the angular velocity increases. At larger angular velocities, the gap between curves corresponding to different $\omega$ values increases significantly, signaling an emergent sensitivity of the electric charge chemical potential to further increases in rotation. 
The effect of the strangeness neutrality condition is directly encoded in the leading-order coefficient $s_1 \equiv s_1(T,\,\omega) = (\mu_S/\mu_B)_{\mathrm{LO}}$ so as to arrive at Eq.~(\ref{s1}) and is shown in the right panel of Fig.~\ref{figure4}. It presents the temperature dependence of $s_1$ at different rotational strengths. In contrast to the electric charge sector characterized by $q_1$, the coefficient $s_1$ takes positive values when $\omega$ = 0. With the inclusion of rotation, $s_1$ increases further, reflecting a similar underlying mechanism to the enhancement observed in $-q_1$.

These LO coefficients can then be used to determine the corresponding chemical potentials by using $\mu_Q \approx q_1\mu_B$ and $\mu_S \approx s_1\mu_B$, which we refer here as method 2. The resulting values are then compared with those obtained by imposing a fixed charge-to-baryon ratio and strangeness neutrality (method-1) in Eq.~(\ref{eq:freeze2}) and (\ref{eq:freeze3}).
In the left panel of Fig.~\ref{figure5}, both the $\mu_Q$ and $\mu_S$ are shown as functions of the baryon chemical potential $\mu_B$ at a fixed temperature $T$ = 160 MeV in the non-rotating limit ($\omega = 0$), and compared for the two methods. 
In the regime of low $\mu_B$, where higher-order contributions are insignificant, the LO approximations from method-2 are expected to hold and to be consistent with those from method-1. This is confirmed in the figure, where both approaches yield nearly identical results at low baryochemical potential, up to $\mu_B$ $\sim$ 100 MeV. However, as $\mu_B$ increases, deviations between the two methods become progressively more pronounced, indicating the growing importance of higher-order contributions. Consequently, the inclusion of NLO coefficients becomes necessary at larger $\mu_B$.
We then proceed to study the effect of rotation by analyzing the variation of these chemical potentials as functions of the angular velocity $\omega$. The right panel of Fig.~\ref{figure5} shows the results obtained using both methods while keeping the temperature and baryon chemical potential fixed at $T = 160$ MeV and $\mu_B = 200$ MeV, respectively. 
Our results show that rotation modifies these chemical potentials: the electric-charge chemical potential $\mu_Q$ decreases with increasing $\omega$, while the strangeness chemical potential $\mu_S$ increases. The difference between the two methods remains constant throughout the ranges of $\omega$ studied for the given $\mu_B$.

\begin{figure*}[ht!]
\begin{center}
\includegraphics[scale = 0.5]{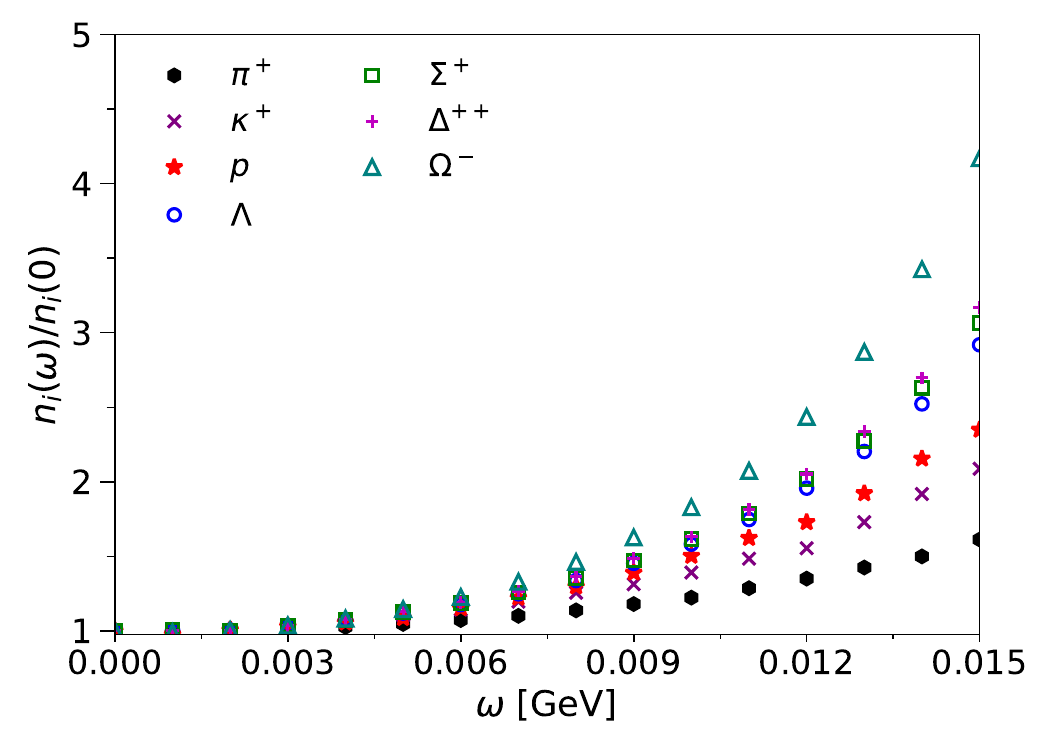}
\includegraphics[scale = 0.5]{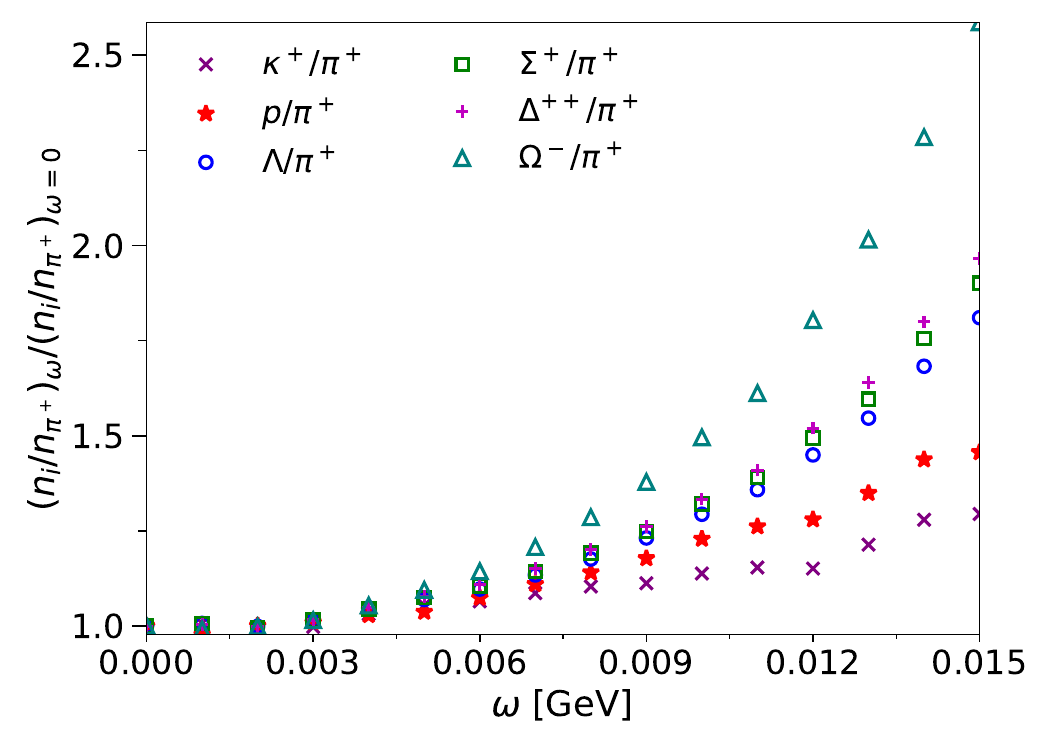}
\caption{(Colour Online) Particle densities (left) and hadron-to-pion densities ratios (right) normalized to their values at zero rotation at vanishing baryochemical potential and $T = 0.155$~GeV are shown as a function of $\omega$.}
\label{figure6}
\end{center}
\end{figure*}

The significant influence of rotation on the freeze-out parameters indicates that its effects cannot be neglected when extracting these parameters from experimental data, particularly in the high $\mu_B$ region corresponding to low-energy heavy-ion collisions. Peripheral collisions are known to generate both strong magnetic fields and large vorticity in the medium. While these effects play an important role in the evolution dynamics of the system and consequently impact the final hadron yields, their experimental quantification remains challenging. Recent HRG model studies~\cite{Fukushima:2016vix} suggest that the enhancement of electric charge susceptibility can serve as a sensitive probe of magnetic fields in heavy-ion collisions. In addition, lQCD calculations~\cite{Bazavov:2012vg, Ding:2025jfz} have proposed the baryon number–electric charge correlation, $\chi_{11}^{BQ}$, as a potential magnetometer of QCD. Furthermore, it has been shown in Ref.~\cite{Vovchenko:2024wbg} that modifications in hadron yield ratios, such as the enhancement of the proton-to-pion ratio ($p/\pi$) and the suppression of the neutron-to-pion ratio, can also act as indicators of magnetic field effects. These observations motivate the search for analogous observables sensitive to rotation to estimate the magnitude of vorticity produced in heavy-ion collisions. With this in mind, we first investigate the influence of rotation on hadronic yields and susceptibilities. 
In Ref.~\cite{Sarma:2025caj}, an attempt was made to understand the effect of both rotation and magnetic fields on hadron yields.
In the present work, we treat rotation as an external field and examine its influence on particle yields and susceptibility ratios to identify potential observables particularly sensitive to rotational effects. In the left panel of Fig.~\ref{figure6}, we show the number density of different particle species normalized by their corresponding values in the absence of rotation as a function of $\omega$ at $\mu_B$ = 0 and $T$ = 155 MeV. In analogy with the magnetic field case, rotation modifies the number densities of hadrons through its coupling to the angular momentum of the particles. In a rotating medium, this interaction induces a shift in the single-particle energy spectrum, which in turn alters the phase-space distribution functions. As a result, the number density acquires a rotation-dependent correction, as expressed in Eq.~(\ref{eq_numden}). From a physical standpoint, this energy shift effectively redistributes the hadron populations, leading to modifications in the thermodynamic observables as well as the corresponding chemical potentials.
\begin{figure}[ht!]
\begin{center}
\includegraphics[scale = 0.53]{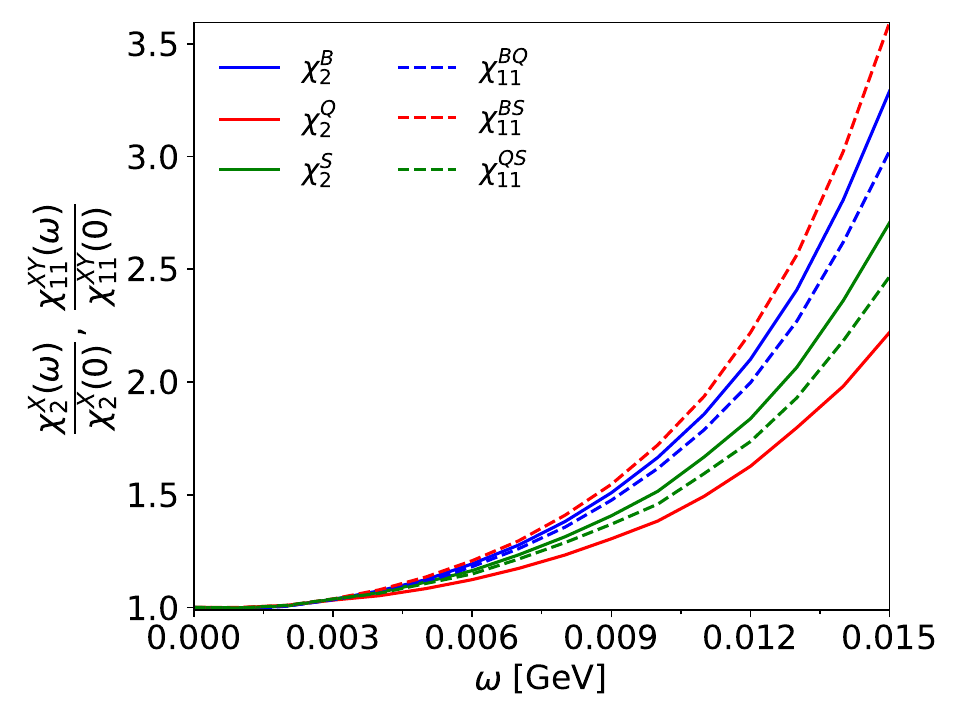}
\caption{(Color online) Ratios of $\chi_{2}^{X}$ and $\chi_{11}^{XY}$ to their corresponding values at zero rotation as functions of $\omega$, at $T = 155\,\mathrm{MeV}$ and $\mu_B = 0$, obtained within the HRG model. Here, $X, Y = B, Q, S$ denote conserved charges. $\chi_{2}^{X}$ represents diagonal susceptibilities, while $\chi_{11}^{XY}$ corresponds to off-diagonal susceptibilities.}
\label{figure7}
\end{center}
\end{figure}

\begin{figure*}[ht!]
\begin{center}
\includegraphics[scale = 0.36]{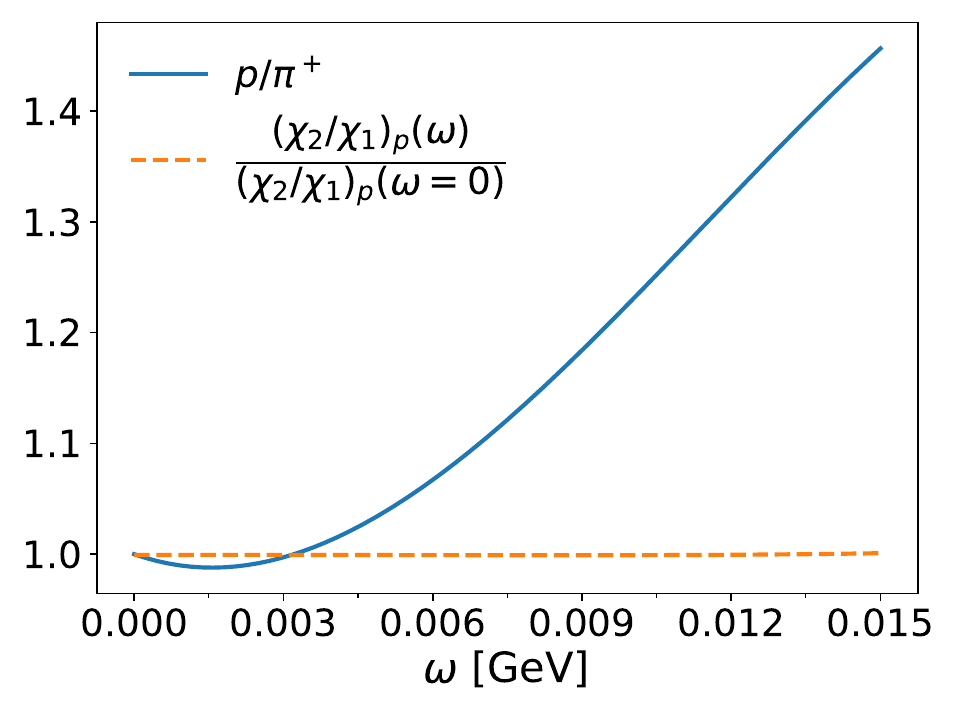}
\includegraphics[scale = 0.36]{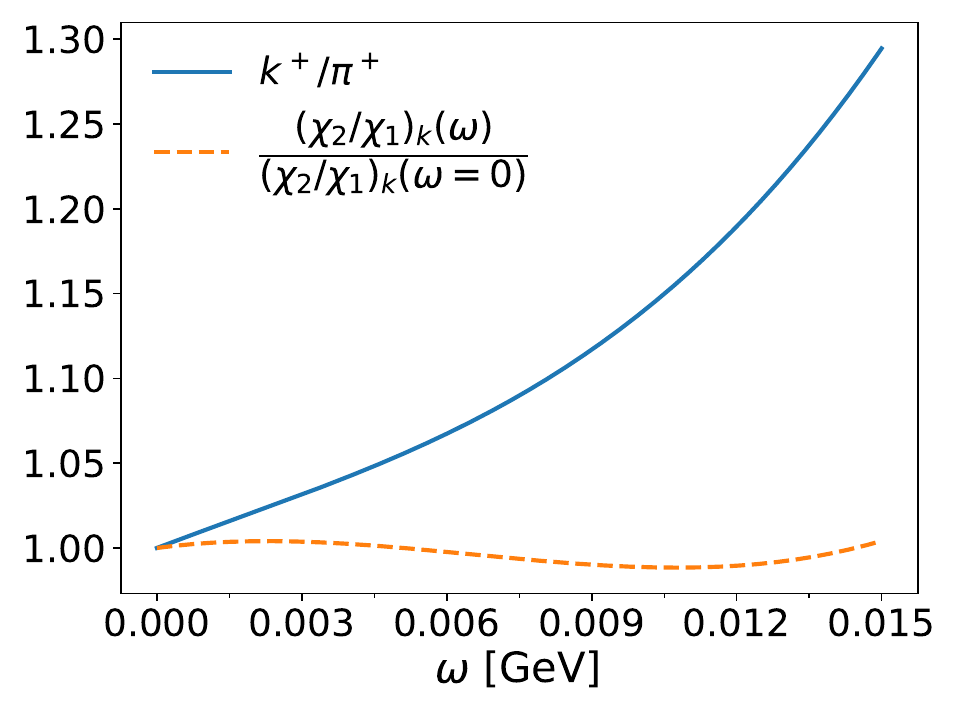}
\includegraphics[scale = 0.36]{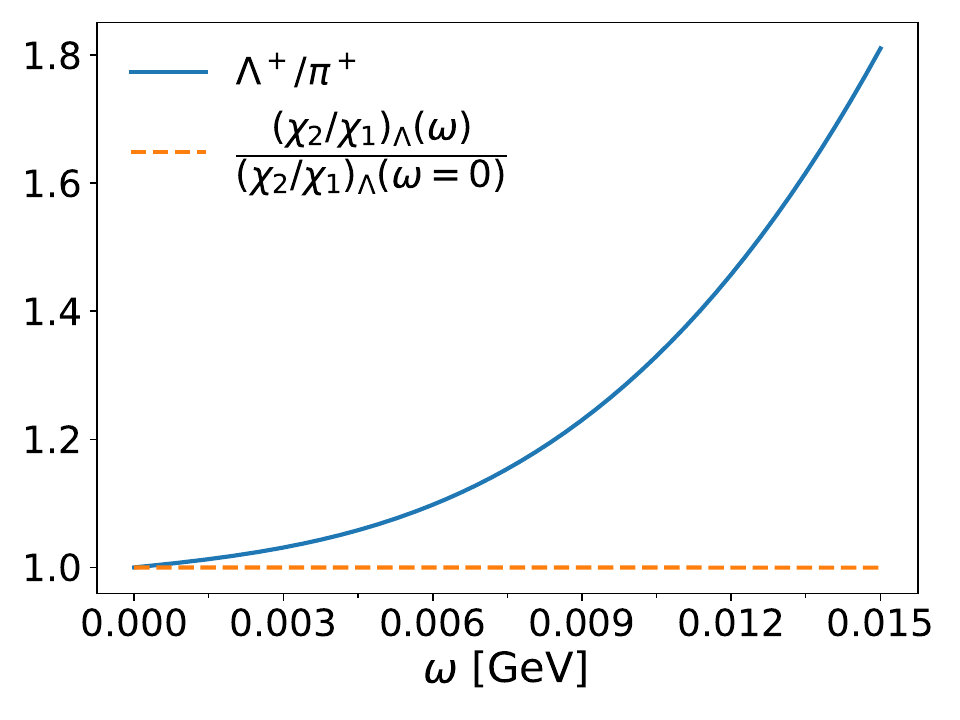}
\caption{(Colour Online) Comparison of the sensitivity of the $\chi_2/\chi_1$ ratio and particle yield ratios to rotation, normalized to their respective values at zero rotation.}
\label{figure8}
\end{center}
\end{figure*}
One can observe a significant enhancement of the $\Omega^{-}$ yield in the presence of rotation, which can be attributed to its large spin ($S = 3/2$). A clear mass ordering is evident across particles ranging from spin-0 to spin-3/2, with heavier particles exhibiting a stronger response to rotation. For instance, the ratio is larger for $\Omega^{-}$ than for $\Delta^{++}$. Although both baryons have the same spin, the zero-rotation number density $n_i$ (0) in the denominator is smaller for $\Omega^{-}$ compared to $\Delta^{++}$, and this difference is not sufficiently compensated by the corresponding value of $n_i$ ($\omega$). This leads to a larger ratio for $\Omega^{-}$, indicating that rotation preferentially enhances the production of heavier hadrons. In the literature, experimental particle yields are commonly expressed as ratios rather than absolute values to eliminate the dependence on the system size. Accordingly, to enable a direct comparison of particle ratios, we present the yields normalized to those of $\pi^+$ in the right panel of Fig.~\ref{figure6}. It displays the ratio of the number densities of various particles to that of pions in finite rotation, normalised with those at zero rotation, as a function of $\omega$. While considering ratios with respect to pions, the effect of rotation becomes clearer, as pions (being spin-zero particles) are comparatively less affected. One can observe that as one moves towards high $\omega$ values, the strongest effect appears in the $\Omega^-/\pi^+$ ratio, followed by the  $\Delta^{++}/\pi^+$, $\Sigma^+/\pi^+$, $\Lambda/\pi^+$, $p/\pi^+$, $k^+/\pi^+$ ratios. Therefore, the $\Omega^-/\pi^+$ ratio is expected to be the most sensitive to the presence of nonzero rotation.
Owing to its large mass and high spin, the $\Omega^-$ hyperon acts as a more sensitive probe of the medium's vorticity than the other hadrons studied here.

Furthermore, we investigate the effect of rotation on the fluctuations and correlations of baryon number, electric charge, and strangeness quantum numbers. These quantities can serve as probes to study changes in degrees of freedom and the QCD phase structure~\cite{Pradhan:2023etz, Stephanov:1998dy, Stephanov:1999zu}, and can also be used to extract the magnitude of rotation by comparing the experimental observations. The susceptibilities $\chi_{2}^{B}$, $\chi_{2}^{Q}$, and $\chi_{2}^{S}$ receive contributions from all baryons (charged and neutral), all electrically charged hadrons, and all strange hadrons, respectively, while the mixed correlations $\chi_{11}^{BQ}$, $\chi_{11}^{BS}$, and $\chi_{11}^{QS}$ arise only from hadrons that simultaneously carry the corresponding conserved charges.
All quantities are normalized to their corresponding values at $\omega = 0$ and are presented as functions of $\omega$ in Fig.~\ref{figure7}. A similar qualitative trend with rotation is observed across all fluctuations and correlations, with differences only in their magnitude. Among these, $\chi_2^Q$ shows the least sensitivity, while $\chi_{11}^{BS}$ exhibits the strongest response to rotation. This enhanced sensitivity of $\chi_{11}^{BS}$ can be attributed to the contribution of heavier, higher-spin strange baryons, which are more significantly affected by rotational effects.

In previous studies, freeze-out parameters have been extracted from experimental data within the framework of statistical hadronization models using hadron yield ratios~\cite{Cleymans:2005xv}. In addition, susceptibility or cumulant ratios of conserved charges have also been employed to determine these parameters~\cite{Alba:2014eba}. To assess the relative sensitivity of these approaches, the authors in Ref.~\cite{Alba:2015iva} compared particle yield ratios (normalized to pion yields) with the lowest-order cumulant ratios and demonstrated their differing sensitivity to the freeze-out temperature. They found that for most hadron species, such as protons, both observables exhibit comparable sensitivity to the temperature. However, for certain species, such as kaons, the net-kaon cumulant ratio ($\chi_2/\chi_1$) provides a more sensitive and reliable probe of the freeze-out temperature compared to the $k/\pi$ yield ratio. 
In the same line, a comparative analysis of the sensitivity to rotation for particle yield ratios (normalized to the pion yield) and the lowest-order susceptibility ratio ($\chi_2/\chi_1$) across some of the hadron species ($p, k, \Lambda$) is shown in Fig.~\ref{figure8}. As discussed in Ref.~\cite{Alba:2015iva}, the lowest-order cumulant ratios are employed, as higher-order cumulants can be significantly influenced by non-statistical effects such as volume fluctuations and critical dynamics. Here, all observables are normalized to their respective values at $\omega = 0$. It is observed that particle yield ratios exhibit a much stronger dependence on the rotation parameter $\omega$, whereas the cumulant ratio $\chi_2/\chi_1$ remains less sensitive, showing only minimal variation over the same range of $\omega$. It is noteworthy that although the susceptibilities are significantly enhanced in the presence of rotation, as shown in Fig.~\ref{figure7}, their ratios, particularly lower-order ones such as $\chi_2/\chi_1$, exhibit only weak sensitivity to rotation. While cumulant ratios are generally preferred because they eliminate the explicit dependence on system volume and reduce experimental uncertainties, enabling a direct comparison between theoretical calculations and experimental measurements, they are found to be less sensitive for probing rotational effects in the medium. This distinct behavior suggests that particle yield ratios are more sensitive to rotational effects in the medium and may therefore serve as a more robust observable for probing and extracting the magnitude of rotation in heavy-ion collisions. In contrast, the weak response of the cumulant ratio indicates its limited effectiveness for such studies. We note that the contributions from heavier resonance decays may further modify these results and potentially enhance the observed effects, although the relative sensitivity is expected to remain similar. However, as a first approximation, such contributions are neglected in the present analysis due to the significant computational expense required to calculate them. In a recent study~\cite{Sahoo:2026lrw}, the transverse momentum ($p_{\rm T}$) spectra of various hadrons were analyzed using a Tsallis non-extensive distribution to estimate the vorticity generated in heavy-ion collisions across different energies. The extracted values are broadly consistent with those obtained from hyperon polarization measurements at the STAR experiment~\cite{STAR:2017ckg}, although they carry sizable uncertainties. In this context, the hadronic yield observables proposed in the present work may provide an improved and complementary approach for estimating the magnitude of vorticity in heavy-ion collision experiments.

\section{Conclusion}
\label{sce_summary}

In this work, we investigate the effect of rotation on the chemical freeze-out curve in ultra-relativistic heavy-ion collisions within the framework of the hadron resonance gas (HRG) model. By incorporating rotational effects through the modification of single-particle energies, we extend the conventional freeze-out analysis to a rotating medium and examine its implications for thermodynamic observables and experimentally relevant quantities. 
We summarize our findings as follows:

\begin{itemize}

\item We observe that rotation leads to a downward shift of the freeze-out temperature in the \(T\text{--}\mu_B\) plane using the two commonly employed freeze-out criteria: the constant average energy per particle, \( \varepsilon / n \approx 1.08~\mathrm{GeV} \), and the normalized entropy density, \( s/T^3 \approx 7 \).

\item The shift in the freeze-out temperature shows a non-linear dependence on \(\omega\), which becomes more pronounced at higher \(\mu_B\). This indicates a strong interplay between rotation and baryon density.

\item The chemical potentials \(\mu_Q\) and \(\mu_S\), constrained by the conservation of electric charge and strangeness neutrality, exhibit a significant dependency on 
rotation, particularly at higher angular velocities.

\item Comparison between the direct implementation of conservation of charge constraints (method 1) and the extraction of leading-order coefficients from susceptibilities (method 2) for determining the chemical potentials \(\mu_Q\) and \(\mu_S\) shows good agreement at low \(\mu_B\). However, noticeable deviations at higher $\mu_B$ highlight the importance of higher-order contributions.

\item Rotation affects hadronic observables, such as particle yields, which show a clear dependence on the mass and spin of hadrons. Heavier and higher-spin hadrons, such as \(\Omega^{-}\), exhibit enhanced sensitivity to rotation and can serve as effective probes to rotational effects.

\item Fluctuations and correlations of conserved charges are also influenced by rotation. The \(\chi_{11}^{BS}\) shows the strongest sensitivity due to contributions from strange baryons, while \(\chi_2^Q\) remains comparatively less affected.

\item  A comparison between hadronic yield ratios and cumulant ratios shows that yield ratios are significantly more sensitive to rotation than low-order cumulant ratios. This suggests that hadronic yield ratios can serve as a more effective probe for determining the magnitude of vorticity in heavy-ion collisions.

\end{itemize}
\section*{Acknowledgement}
N.P. acknowledges the financial support from the Ministry of Education (MoE), Government of India. The authors sincerely acknowledge the computational facilities provided by the National Institute of Technology Durgapur.
K.K.P. and R.S. gratefully acknowledge the DAE-DST, Govt. of India funding under the mega-science project -- “Indian participation in the ALICE experiment at CERN" bearing Project No. SR/MF/PS-02/2021-IITI (E-37123).

\end{document}